International Journal of Computer Networks & Communications (IJCNC), Vol.2, No.3, May 2010# AN ENVISION OF LOW COST MOBILE ADHOC NETWORK TEST BED IN A LABORATORY ENVIRONMENT EMULATING AN ACTUAL MANET

Nitiket N Mhala[1] and N K Choudhari [2]

[1]Assistant Professor, Department of Electronics Engg., BDCOE, Sevagram,India
*nitiket_m@rediffmail.com*
[2]Principal, Bhagwati Chadurvedi COE, Nagpur,India
*drnitinchoudhari@gmail.com*
*ABSTRACT*

*Orchestrating a live field trial of wireless mobile networking involves significant cost and logistical issues relating to mobile platforms, support personnel, network and experiment automation and support equipment. The significant cost and logistics required to execute such a field trial can also be limiting in terms of achieving meaningful test results that exercise a practical number of mobile nodes over a significant set of test conditions within a given time. There is no argument that field trials are an important component of dynamic network testing. A field test of prototype will show whether simulations were on right track or not, but that's a big leap to take; going from the simulator directly to the real thing. In conceiving our work, we envisioned a mobile network emulation system that is low cost, flexible and controllable. This paper describes our wireless MANET test bed under development which emulates an actual MANET. Here, we focuses that, this test bed allows the users to automatically generate arbitrary logically network topologies in order to perform real time operations on adhoc network at a relatively low cost in a laboratory environment without having to physically move the nodes in the adhoc network. Thus, we try to "compress" wireless network so that it fits on a single table.*

## 1. INTRODUCTION

Recent technical publications are inundated with reports of promising technologies and approaches for better wireless future. One emerging technology are, mobile adhoc networks (MANET), focuses on support for autonomous operation of dynamic, wireless networks within designated localized areas of a given IP- based wireless networks. This technology is seen as an important advancement for the future of network-centric military operations and has many commercial and emergency-related applications such as vehicle safety, sensor network, first responder assistance, smart homes and factory automation application.  Thus the mobile networking is a hot topic of present technology of research hardly needs saying. Therefore these scenarios have remained largely the domain of university researchers or government funded laboratories. An adhoc network consists of mobile nodes without the required intervention of a centralized access point or existing infrastructure. The links of the network are dynamic and are

10.5121/ijcnc.2010.2305          52

based on the proximity of one node to another node. The links are likely break and change as the node move across the network. Because of the temporary nature of the network links, conventional routing protocols are not appropriate for adhoc mobile networks. Thus adhoc networks are very important part of mobility management.Adhoc networking has to be integrated into the existing infrastructure in future.

## 2. Related Work

Previous Test-Bed works report on experiments with adhoc. In [1] they built multihop wireless adhoc of 8 nodes including both stationary and mobile nodes and drove around a 700m by 300m site. In [2] they developed choreography script, which reports on 37 physical nodes. Testers walk around the building with wireless LAN card equipped laptops. These Test-Bed shows reproducible and scalable experiments, but costs on devices and human resources are very high. In [3] Recreating realistic environmental conditions and signal transmission characteristics using of the shelf computing nodes and wireless cards in a laboratory setting is also very difficult. In [4] they perform Outdoor field tests of mobile adhoc networks reveal real realistic use case scenarios and can validate expected results under real world conditions, but these test can be prohibitively expensive and out of reach for the typical researcher or developer. In [5] they successfully demonstrated realistic feature of mobility but the test bed does not allow a gradual movement of nodes. Similarly there is an abundance of research based on simulations, actual implementation of ad hoc routing protocols and applications are very limited by comparison. Many researchers designed new MANET algorithm / protocol / application / whatever and successfully evaluated its efficiency with extensive simulations. What is the next step? A field test of prototype will show whether simulations were on the right track or not, but that's a big leap to take; going from the simulator directly to the real thing. Despite the profusion of research based on simulation, current attention is inclined towards real world implementation [6] that is why there are test beds to bridge that gap. Test beds are emulators, typically a great number of devices with wireless capabilities deployed in a large space and waiting for the user to program them, experiment with them and test the work almost realistic conditions. However there are few who can afford to have such facility in their laboratory. Usually, it is possible to rent this equipment, run the experiment remotely and then receives the results and analyzes them. The problem is that work will be conducted under serious time constraints, especially when we are on a tight budget. That's where our Test bed comes in. It is also MANET emulator, a test bed, but the difference with other test bed is that it is intended to operate in a laboratory.

### 2. 1 CHALLENGES IN A LABORATORY ENVIRONMENT

Testing adhoc network in a laboratory environment presents a number of challenges. The most obvious challenge is being able to test the effects of node mobility on the adhoc routing protocols and adhoc applications. Moreover, configuring individual nodes, installing patches, monitoring log files, updating software and debugging beta releases of experimental software distributions on a modest size of adhoc network can be very time consuming. Recreating realistic environmental conditions and signal transmission characteristics using off-the-shelf computing nodes and wireless cards in a laboratory setting is also very difficult.

### 3. BASIC IDEA AND RESEARCH METHODOLOGY

Wireless connectivity is the state of the art for local area network. In order to test adhoc network implementations in a laboratory environment, Wireless adhoc network test bed is under development. The Software modules developed to create logical topologies capture the packets transmitted over the wireless interface and capable of providing diagnostic information on packet sent and received by nodes in adhoc network. A really important concept regarding our Test Bed is that it emulates a MANET at the network layer of OSI (layer 3) and above. That means it has nothing to do with neither the physical nor the data link layer(layers 1 and 2.In fact, the nodes will most probably, be close to each other in a laboratory, so every node can "hear" every packet transmitted by any node, but ignores most of them, according to what our test bed dictates. Thus, our test bed does not emulate a MANET at the data link layer; it emulates MANET at the network layer.

The basic idea of our emulation Test Bed is having a number of MANET nodes physically close to each other inside the laboratory, but forces them to 'think' that they can only communicate with a selected few of them. That way, we can emulate a logical topology, on which can run e.g a routing protocol to analyze its behavior. In order to work, there is the need of hardware. Our Test bed is an emulator, not a simulator. So, we have the necessary hardware equipment. Each node is a device that has a wireless (802.11 b/g) interface, so that it can communicate with other adhoc nodes and run MANET protocols. In addition, the device should also have a wired interface (Ethernet), which is used for administrative purposes. In brief, the Test Bed uses the wired interface to transfer files needed for its operation to and from the node and manipulate its networking element in such way that will create logical topology we want. That Leaves the wireless interface free of any interface and most importantly, emulates an actual MANET, which is the whole point all along.

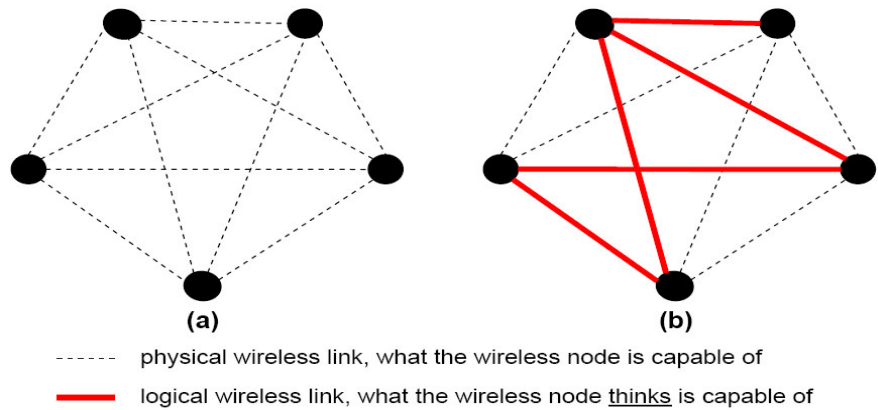

Figure 1. (a) Physical links, (b) logical links between MANET nodes

## 4. REQUIREMENTS OF OUR TEST BED

The hardware platform consist of a central Linux desk top or laptop machine that communicates with member nodes of adhoc network through wired interface while the adhoc nodes communicate with each other trough wireless interface. In fact, our Test Bed uses ix86 architecture and Linux can run even in 80386 machines (at least requirement is Pentium II), so we can gather all those old PCs intended thrown away, adding a PCMCIA wireless card on each of them and set up a MANET test bed in a laboratory at a very low cost.

### 4.1 Laptops/Desktops

At least three nodes with Intel Pentium or higher
Server:  Linux system (ix86) with one wired and one wireless interface
Clients: Any Linux system with one wired and one wireless interface

### 4.2 Wireless LAN Card

Test bed needs Wireless LAN cards (IEEE 802.11) which should be configured to Adhoc mode. Our Choice for Test Bed is Netgear  WG511T 108 Mbps Wireless PC Card and   WG311T 108 Mbps Wireless PCI Card (With Chip set Atheros 5212)
### 4.3 Drivers

Madwifi Driver (madwifi-0.9.4)

A Linux kernel driver for Atheros –based wireless LAN devices. The driver support ad hoc mode of operation [7]

### 4.4 Operating System

Test Bed runs on Fedora Core (FC6/FC7)

### 4.5 Netfilter

Netfilter is the framework in Linux 2.6 kernels that allows for packet filtering and NAT. Iptables is the user space tool that works with Netfilter frame work which is done on the mac-layer

### 4.6 Sed,fgrep,xargs, console

In standard installation of Linux system

### 4.7 Expect (expect-5.43)

Performs programmed dialogue with other interactive program [8]

### 4.8 Libpcap0.9.4 or higher

Libpcap provides a portable framework for low – level monitoring [9]

### 4.9 QT3 (with thread and gif support)

Multiplatform C++ GUI application framework [10]

### 4.10 Greaphviz

Graph Visualization Tool

## 5. TEST BED MODULES

### 5.1 Network Configuration Module

In this module, we can declare MANET nodes to the test bed and transfer the required files to nodes. Addition, Edition and Deletion of the nodes can be done effectively. Each node's wired and wireless IP and MAC address is installed in this module. This configuration can support up to 150 nodes. Theoretically, there is no limit to the number of nodes this test bed can handle.

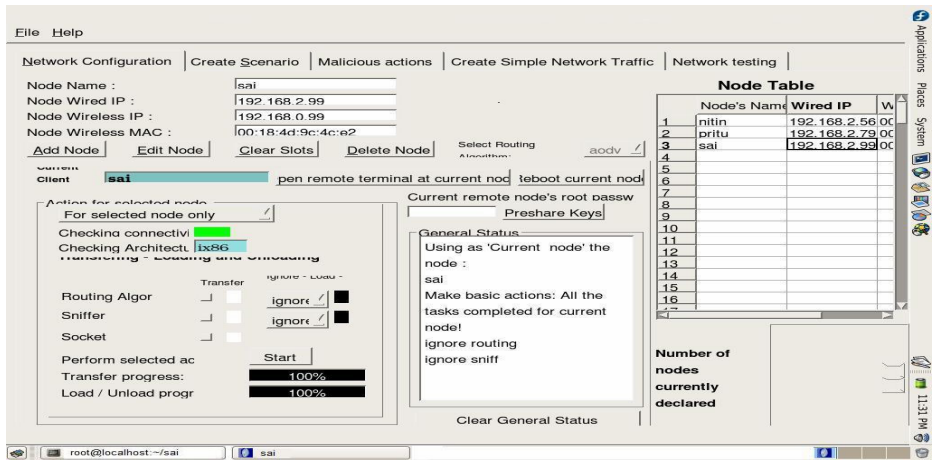

Figure 2. Network Configuration Module

**5.2 Design Parameters for logical Topology creation scenario module**

- Scenario name: It is the name of the scenario to be created by the user in a test bed.
- Number of nodes: How many nodes the generated topologies will have. Of course this should be less or equal to the number of nodes we have declared in the network configuration module
- Number of Topologies: How many topologies this scenario will have
- Maximum node degree per node: The degree of node is how many connections it has with other nodes. When generating random topologies a node will have a random number of connections with other nodes from zero to the value we have set here.
- Network Density (0-100): How dense the network will be. That is determined by an adjacency matrix, where element every element (x, y) determines if there is a connection between node x and node y. The number we have set here is the possibility to have a '1' in each position (x, y), therefore, have connection. And '0' i.e default value search for a node that have a neighbor according to the adjacency matrix. We take a starting node of the adjacency matrix that we have created and then based on that try to see if the adjacency matrixes that we have produced give us a connected graph. That here we discover a starting node and then, based on the neighbours, we try to include all the nodes of the adjacency matrix.

We assume Boolean connectivity between the nodes. This means that two nodes are either connected or they are not.

We set status value as
99 -> the adjacency matrix give us an over connected or disconnected topology
100-> the adjacency matrix give us a suitable topology
Therefore the adjacency matrix is checked to meet our constraints of maximum node degree {1-4} After this point the status value should be 99 (not connected) or 100 (connected)

Thus, this module allows users to save and replay different mobility scenarios, to control the maximum and minimum node degree, produces an output in the form of adjacency matrix for further analysis and provides a framework for building additional ad hoc network testing tools.

**5.3 Topology creation Scenario module**

This scenario is a series of random topologies. After setting various options, several random topologies are generated by the software module. Its purpose is to generate random network topologies, display it to the user using Graphviz tool and to directly apply them (Through sockets at the wired interface) to simulated LAN. Thus the hardware using for the tool is a host-server machine with two network interfaces that will have direct connectivity to the LAN that will be changing the topology and all the client members to that LAN to have two network interfaces.

The users can automatically generate arbitrary logical network topology in order to perform real time performance measurements of routing protocol implementations. By changing the logical topology of the network, users can conduct test on adhoc network without having to physically move the nodes in the adhoc network. Given the number of nodes in adhoc network Test Bed, each node's IP and MAC address, software module used in a test Bed creates arbitrarily connected graphs and updates each node's IP_TABLES accordingly through socket servers running on each network node in order to reflect the new logical topology. Thus arbitrary graph is represented in an adjacency matrix that is then translated into the corresponding IP_TABLES. Software module uses open source graph visualization tool Graphviz [11] to display the logical topology of the adhoc network as shown below.

Figure 3 indicates arbitrary creation of logical topology of physical three nodes in laboratory with manual handling setting to topology number 4 of current scenario 'Topology'.

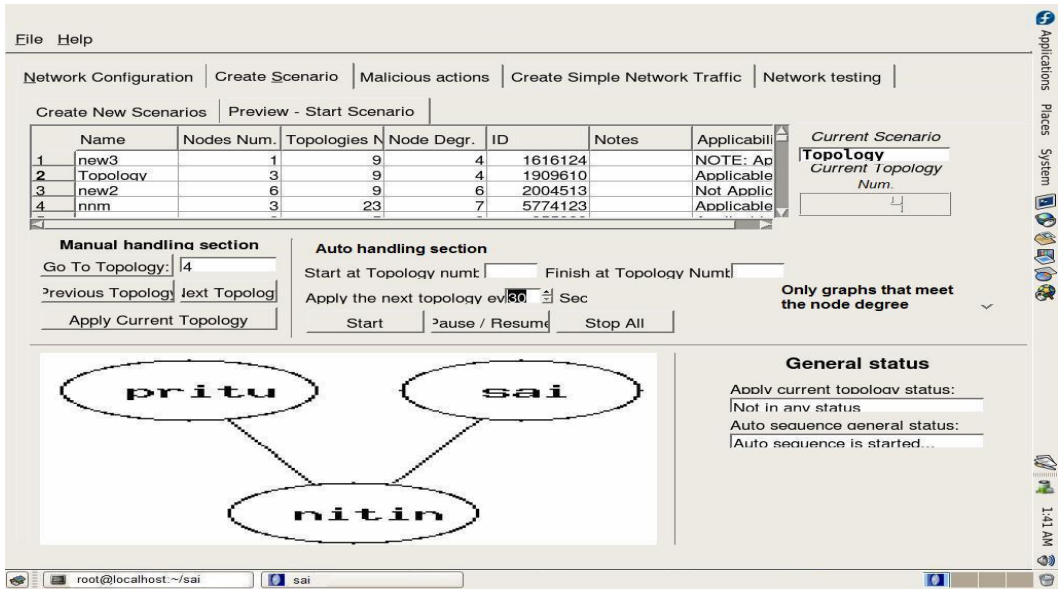

Figure 3. Logical Topology Creation (manual handling)

Figure 4 indicates arbitrary creation of logical topology of three nodes automatically after every 30 seconds for setting to topology number 0 to 9 for the same current scenario 'Topology'.

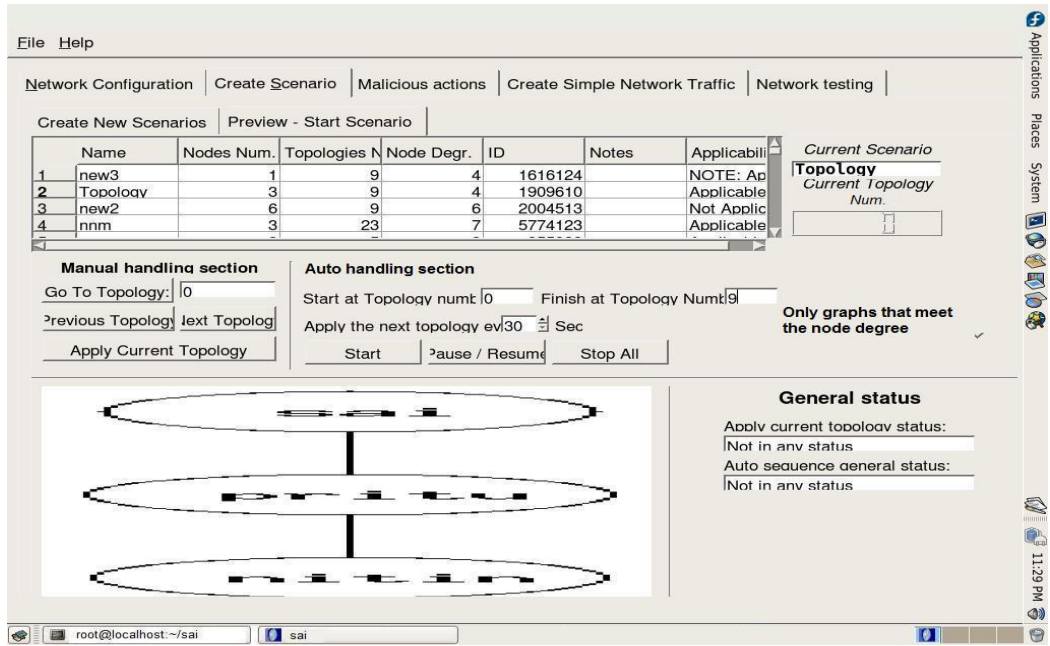

Figure 4. Logical Topology Creation

Figure 5 again indicates arbitrary creation of logical topology of actual three nodes in every 30 seconds for setting to topology number 0 to 9 for the same current scenario 'Topology'.

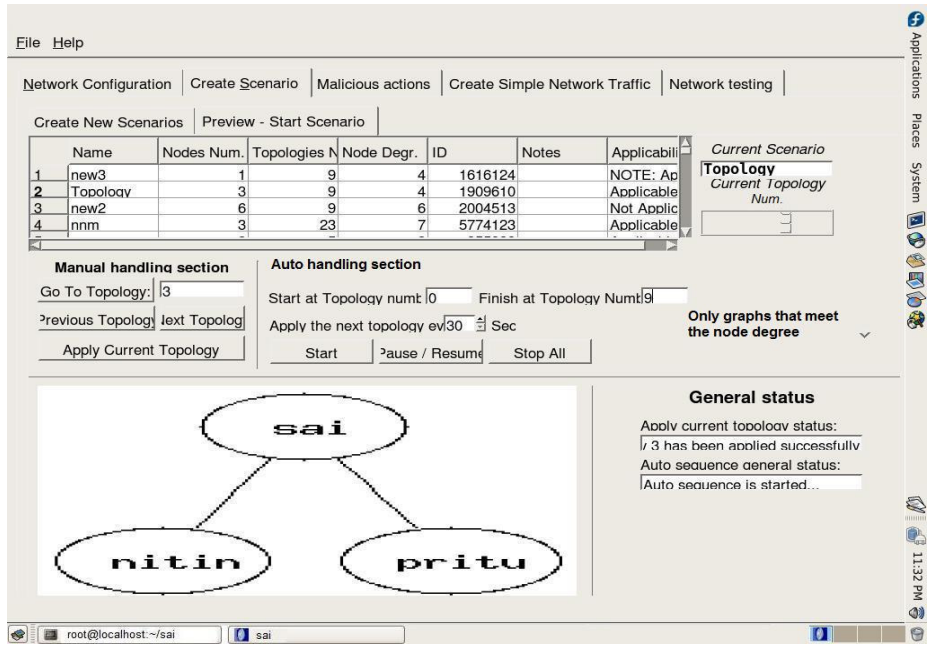

Figure 5. Logical Topology Creation

### 5.4 Malicious Actions Module

In this module we can select from a range of the most common attacks against some of the most common protocols or just inject into the network any packet we want in a hexadecimal form. We can see the logical diagram of our wireless network in this module what we have already created in Topology creation scenario module. We can choose to attack against TCP, UDP or ICMP protocol.

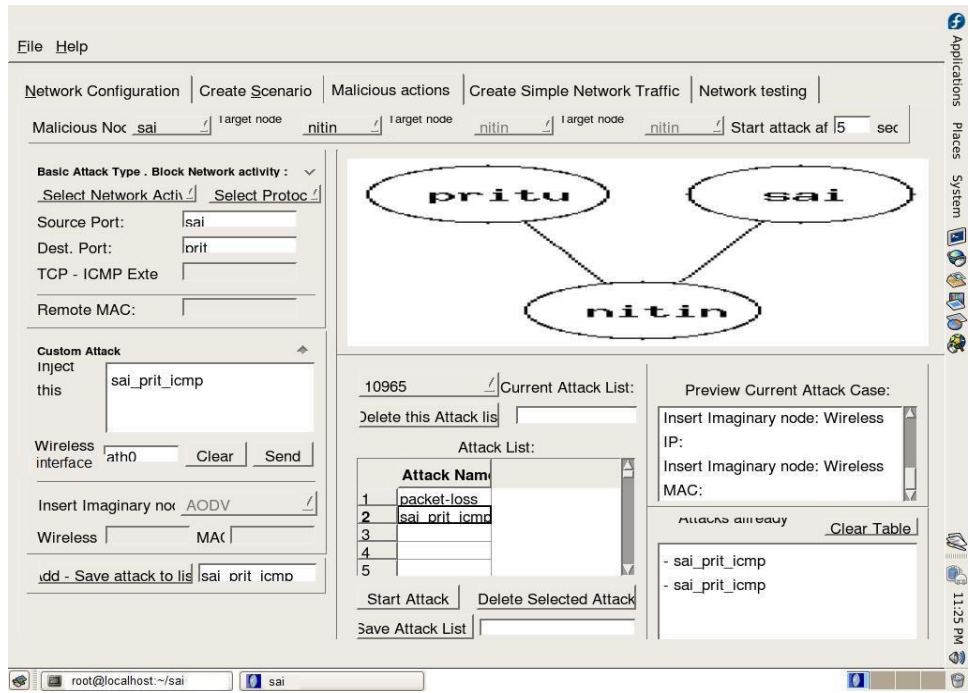

Figure 6. malicious action module

58

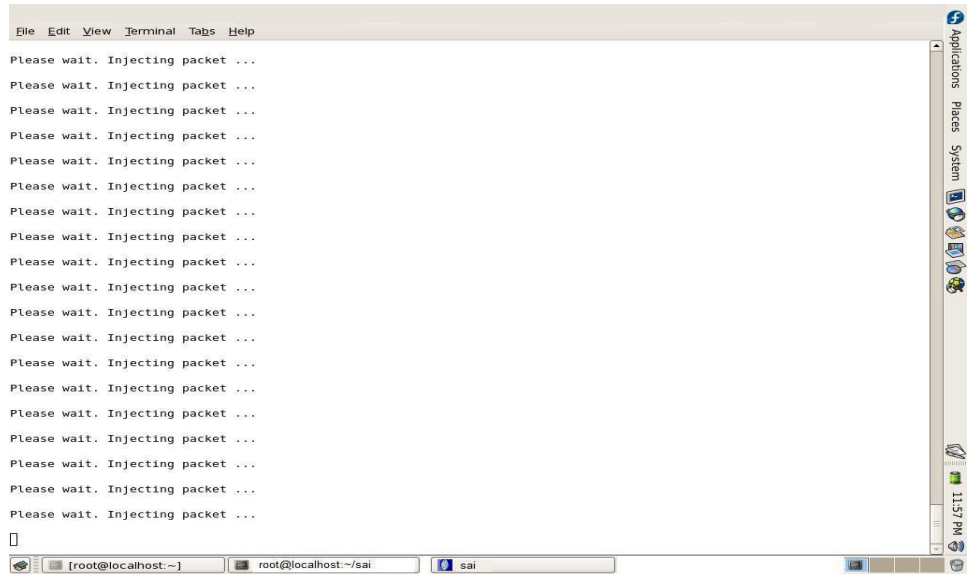

Figure 7. Injecting of packets

The Types of attacks are Blocking all incoming traffic, Blocking all outgoing traffic, Blocking both incoming and outgoing traffic, Generating packet loss.
More specifically, it will generate packet loss for 5 seconds, return to normal for the next 35 seconds and start over again for a total of 10 times. After, spitefully, planning the attack, we can give appropriately evil name and it is saved and added to the attack list in order to repeat it any time.

### 5.5 Network Traffic Generator Module

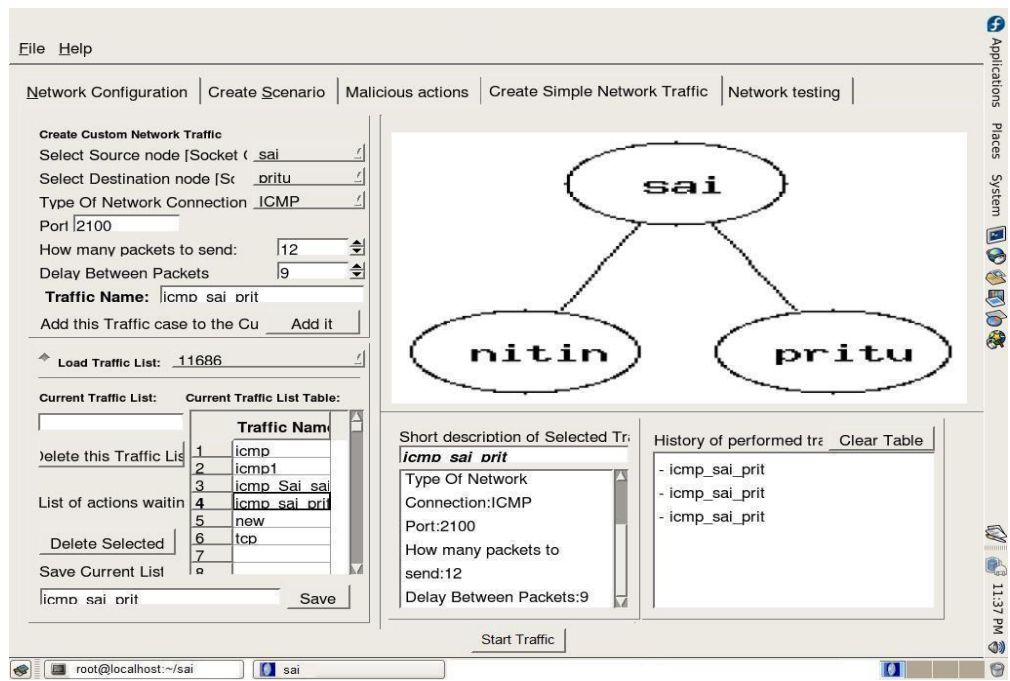

Figure 8. Network traffic generator (ICMP traffic sai to pritu)

Here, we can create a simple, constant data flow between two nodes in the network (peer to peer connection).Again, we can see the logical diagram of the network at the upper right section of the screen. We can select the source and destination node, the protocol (TCP, UDP, ICMP),the port to be used and similarly how long the delay will be between two consecutive paths.

### 5.6 Quick Network Testing Module

It is an important module or tool to check whether or not our new topology case has been successfully applied. Also, we can execute any command line command we wish on any remote node. It is necessary that while the remote command is being executed, the test bed software is being focused on it and no other Test Bed fixture is available during the command execution.

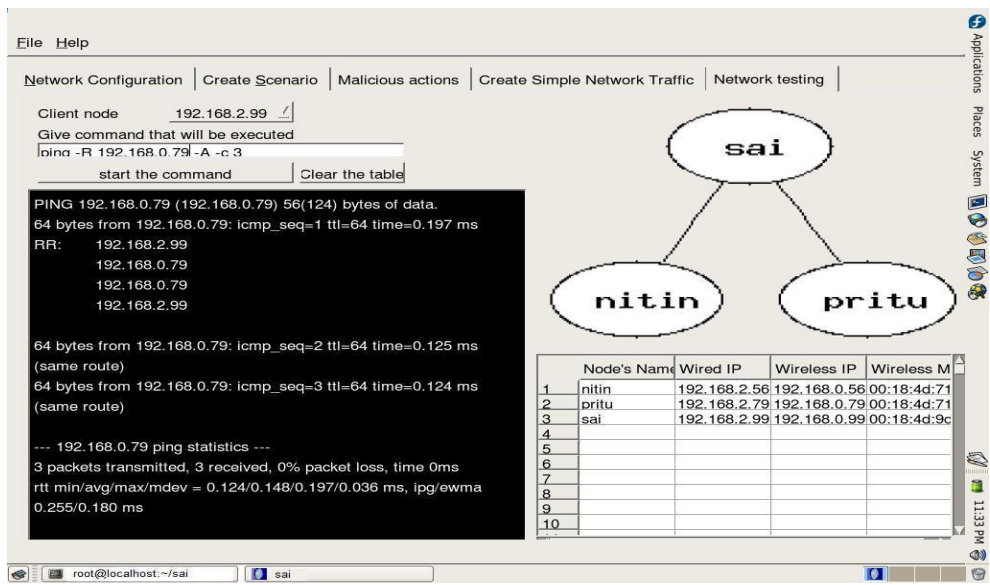

Figure 9

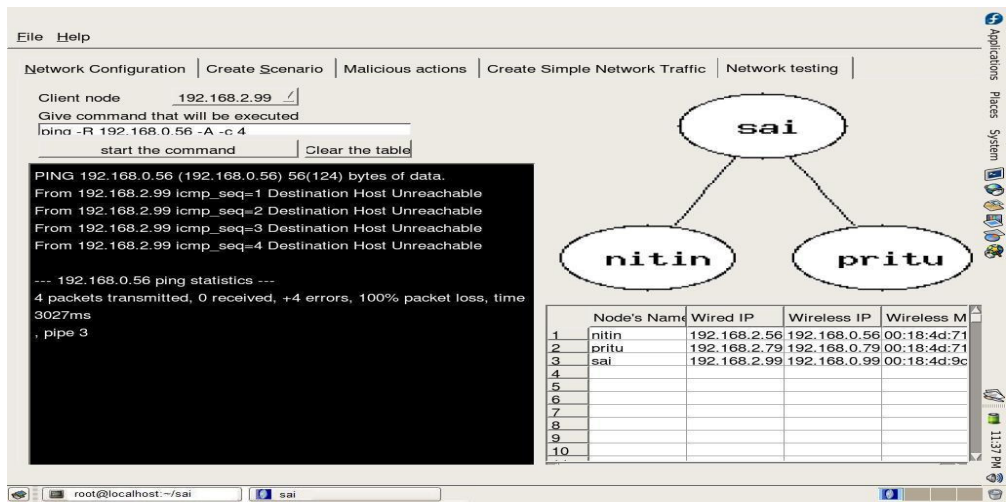

Figure 10

Fig 9 illustrates the network testing for the created logical topology of the network. From node Sai to Node Pritu a connection is tested by use of ping command .Three packets are sent from node sai to node pritu. A connection is tested successfully by receiving three packets by node Pritu showing the statics of 0 % packet loss.

Fig10 illustrate the 100 % packet loss for the node nitin for the same topology as the case illustrated in fig 9.This confirms that Destination node nitin is not reachable.

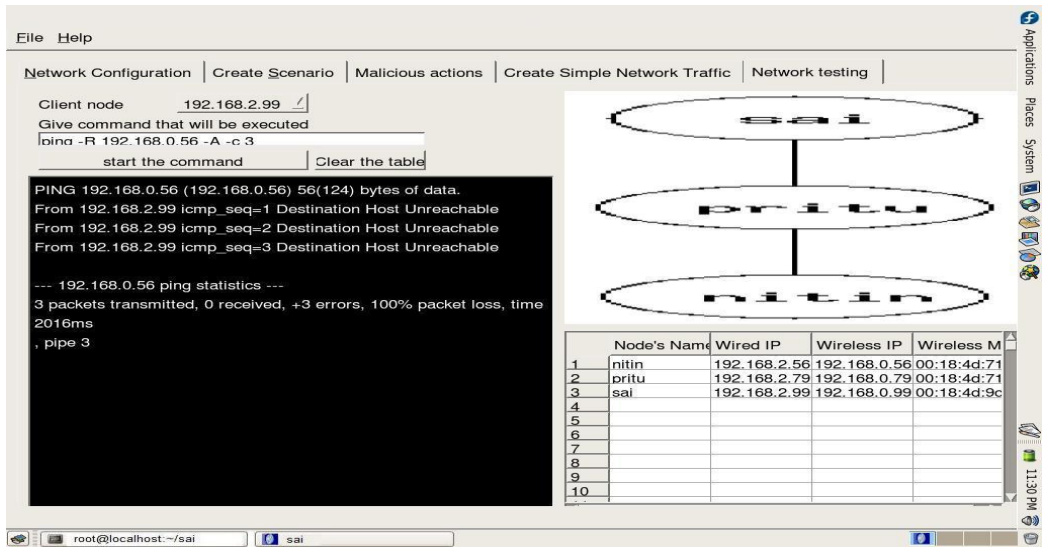

Figure 11

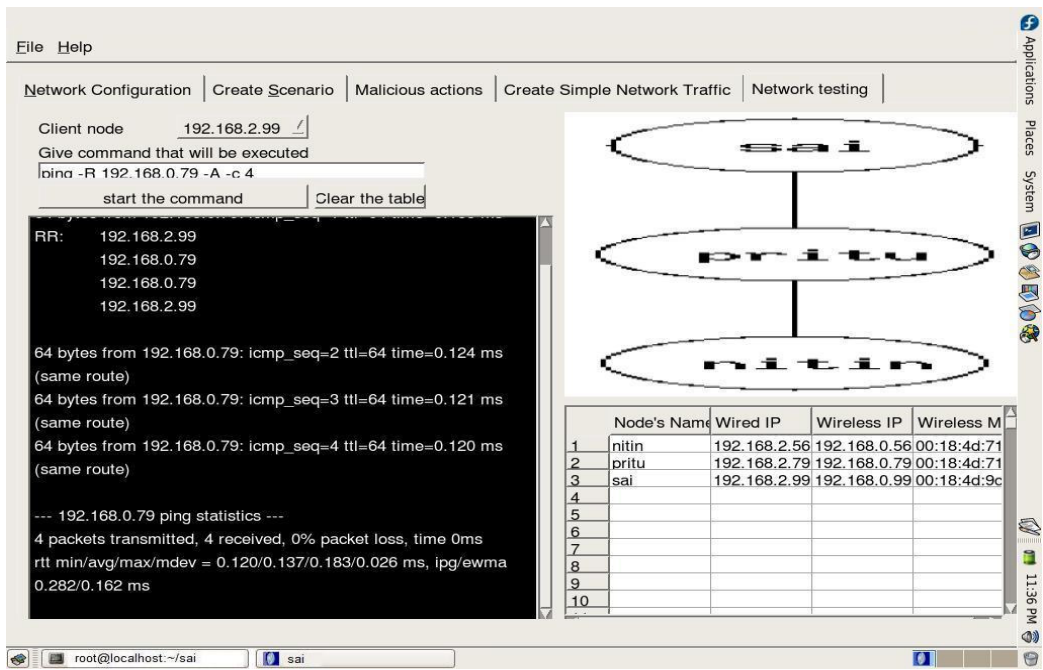

Figure 12

Fig 11 shows different logical topology from where node again nitin is tested. But again the destination node nitin is not reachable having 100% packet loss.

Fig 12 similarly shows for the same topology applied as in case of fig 11, the node Pritu is being tested by the same ping command. The statists confirm 0 % packet loss.

## 6. CONCLUSION AND FUTURE WORK

We have successfully demonstrated a low cost flexible Test Bed as Real MANET emulator at the network layer of OSI (layer 3) and above. Although we are able to control each node's logical perception of the network topology, we cannot control the network congestion resulting from having a large number of adhoc nodes in a close proximity in a relatively small research laboratory. But this issue will not prevent budget-conscious researchers from collecting useful results a modest size network. In fact, determining the global network topology in a mobile adhoc network given the time delays of the diagnostic packets and the mobility of the nodes make this task very difficult, but determining an approximation of this topology or subset of this topology, within certain time frame may be useful [12] .An approximation of the network topology can still provide the useful information about network density, network mobility, critical paths and critical nodes. Even with the uncertainty associated with correctly reconstructing the network topology for a given time period, this additional information can help reduce resource consumed to monitor all nodes in the absence of this information.

In an adhoc network, malicious node may enter and leave the intermediate radio transmission range at random interval, may collude with other malicious nodes to disrupt network activity or behave maliciously. Packets may be dropped due to network congestion or because a malicious node is not faithfully executing a routing algorithm [13]. Our test bed supports the implementation of AODV protocol [14].The Test bed under development provides successive arbitrary graphs at predefined time intervals. When new logical topologies have been generated, the network will experience an increase in the number of RREQ and RERR messages. Therefore, Intrusion Detection system (IDS) must be able to differentiate legitimate RERR messages and those that are the result of malicious activity. Generating arbitrary topologies create scenarios that can help researchers test experimental IDS system under the difficult conditions and this will form the basis for our future work [15] In conclusion, this paper submits that since the global topology of the adhoc network is known, researchers can benchmark the actual performance of their adhoc routing algorithms and applications against the theoretical optimal performance.

## ACKNOWLEDGEMENT

The authors would like to thank everyone, including the anonymous reviewers.

**Authors**

**Mr. Nitiket N. Mhala** is PhD student and also working as Assistant Professor in the Department of Electronic Engineering, Sevagram, India. He received his ME Degree from RM Institute of Research and Technology, Badnera, Amravati University and BE Degree from Govt. College of Engineering, Amravati, Amravati University. He published a Book Entitled PC Architecture and Maintenance. He is a member of Institute of Electronics and Telecommunication Engineer (IETE). His area of interest spans Data communication, Computer network and Wireless Ad hoc networks.

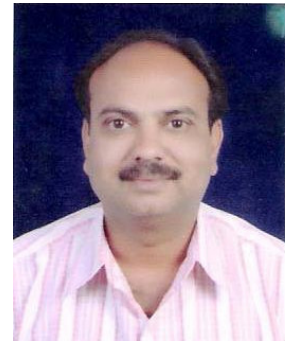

**Dr. N. K. Choudhari** is a Professor and completed his Ph.D degree in Electronics Engineering from J.M.I., New Delhi and received his M.Tech in Electronics Engineering from Visveswaraya regional Engineering College, Nagpur. He received his BE in Power Electronics from B.D.C.O.E., Sevagram. Presently he is Principal at Smt.Bhagwati Chaturvedi COE, Nagpur, India. He is guiding few research scholars for persuing Ph.D degree in RTM Nagpur University, Nagpur, India. He has worked as members of different advisory committees and was a member of Board of Studies Electronics Engg. of RTM Nagpur University, Nagpur, India.

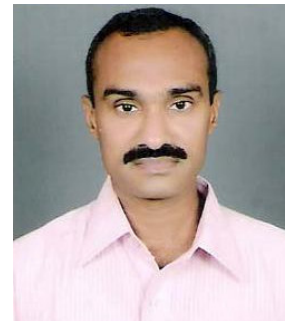